\documentclass{acm_proc_article-sp}
\usepackage{graphicx}

\begin{document}

\conferenceinfo{}{Bloomberg Data for Good Exchange 2016, NY, USA}

\title{Digitizing Municipal Street Inspections Using Computer Vision}

\numberofauthors{4}
\author{
\alignauthor
Varun Adibhatla\\
       \affaddr{ARGO Labs}\\
       \affaddr{New York City, NY}\\
       \email{varun@argolabs.org}
\alignauthor
Shi Fan\\
       \affaddr{NYU Center for Data Science}\\
       \affaddr{New York City, NY}\\
       \email{sf2632@nyu.edu}
\and
\alignauthor Krystof Litomisky\\
       \affaddr{ARGO Labs}\\
       \affaddr{La Cresenta, CA}\\
       \email{krystof@litomisky.com}
\alignauthor Patrick Atwater\\
       \affaddr{ARGO Labs}\\
       \affaddr{La Canada Flintrige, CA}\\
       \email{patrick@argolabs.org}
}

\maketitle
\begin{abstract}
``People want an authority to tell them how to value things. But they chose this authority not based on facts or results. They chose it because it seems authoritative and familiar." - The Big Short [1]

The pavement condition index is one such a familiar measure used by many US cities to measure street quality and justify billions of dollars spent every year on street repair [2]. These billion-dollar decisions are based on evaluation criteria that are subjective and not representative. In this paper, we build upon our initial submission to D4GX 2015[10] that approaches this problem of information asymmetry in municipal decision-making.

We describe a process to identify street-defects using computer vision techniques on data collected using the Street Quality Identification Device (SQUID). A User Interface to host a large quantity of image data towards digitizing the street inspection process and enabling actionable intelligence for a core public service is also described. This approach of combining device, data and decision-making around street repair enables cities make targeted decisions about street repair and could lead to an anticipatory response which can result in significant cost savings. Lastly, we share lessons learnt from the deployment of SQUID in the city of Syracuse, NY.
\end{abstract}





\section{Introduction}
Pavement condition index or PCI is a measure of pavement distress on a scale of 0 to 100, calculated from visual assessment on a sample of road networks [3]. The standard was originally developed in the 1970s by U.S. Army Corps of Engineers. The PCI relies on city employees manually surveying city streets using a prescriptive manual that contains a visual reference of various pavement distresses and street defects. These employees often undergo some training prior to surveying city streets. We argue that regardless of training or expertise, these visual measurements are subject to inconsistencies and error. Furthermore, the PCI is often conducted on a sampling of city streets and not the entire street grid. To that end any repair or maintenance intervention premised on these measurements is not representative and not equitable.

An alternative to the PCI is the International Roughness Index (IRI) involves using laser technology to assess streets at a very high precision. Unfortunately, for most US cities, IRI is not a viable option for citywide street quality measurement due to its high cost [2]. Many US Cities are being challenged to think boldly about their transportation system and prepare for a driverless future [4]. We believe that the same form of bold thinking should apply to the maintenance, repair and response of the core piece of municipal infrastructure, roads.

There is a clear need for a low-cost process that enables city agencies to measure the entire street grid using a data gathering process that is verifiable, repeatable, and actionable to allow for complete and longitudinal measurements of street quality. This will not only allow cities to make more targeted and efficient decisions about street resurfacing, but also enables an anticipatory response paradigm which allows cities to be more responsive and resource and capital efficient around street infrastructure [5].

According to the American Association of State Highway and Transportation Officials, ``every \$1 spent to keep a road in good condition avoids \$6-14 needed later to rebuild the same road once it has deteriorated significantly." [6]

Furthermore, the director of the Los Angeles' Bureau of Street Services echoes a similar concern ``for every block I do reconstruction, I could have done five to seven blocks of resurfacing." [7] The reason is that resurfacing usually entails replacing or installing a new top layer of asphalt pavement, whereas reconstruction ``replaces over a foot of the roadway below the street's surface and usually includes reconstruction of the curbs and sidewalks as well." [8] It is therefore crucial for city governments to make proactive decisions to prevent damaged streets from undergoing reconstruction by resurfacing them in a timely manner.

\newpage
\section{Process Overview}
Digitizing street inspection is not a novel idea by itself. A review of digital techniques to street quality assessments is presented in [10]. However, few attempts solve for scalability. Our approach involves a ``soup to nuts" development cycle that consists of a low-cost device, passive data collection and cloud storage, to automated computer vision-based road surface inspections and public-facing web-based interactive visualization tools. In [11], a similar technique using computer vision and mobile devices is described. In this paper, we build upon these advances in academia and the private sector in delivering a holistic approach that can not only benefit municipal managers but also address systemic issues in citywide street maintenance.

In this section, we provide a brief overview of our product, SQUID, from how we design, assemble and implement the device to how the data gets collected, stored and cleaned in a robust and RESTful manner.

\subsection{Device}
SQUID consists of a Raspberry Pi 2 B computer, a camera, an accelerometer, and a GPS platform. It is a low-cost device, but it is well-suited to digitizing street inspection process. In terms of implementation, the device is mounted to the back of a city vehicle with its camera facing downwards to the road. A more detailed description of the device is offered in [10].

\subsection{Data}
\subsubsection{Collection}
SQUID collects imagery, location and ``ride quality data" from camera, GPS and accelerometer. The data collection frequency is set at 1 Hz (1 image per second) and is done passively. The only constraint is to maintain driving speed to below 35 mph to capture a high quality image and stay within local speed limits. In addition to imagery, location and time, ``ride quality data" collected at each second includes vehicle speed as well as acceleration in all directions (x, y, z).

Ride Quality Score = $\sqrt{x^2 + y^2 + z^2}$

where x,y \& z are measures of acceleration.

\subsubsection{Storage}
After data collection, we release the data out of the device as quickly as possible and avoid performing any complex computation on the device itself. This allows us to leverage inexpensive data storage and elastic computing services while keeping costs and complexity low on the device side. To this end, we use a mobile hotspot to transmit the data we collect in real time to the cloud. Alternatively, we can also leverage strategically located secure municipal wireless internet to serve as an asynchronous uplink. SQUID uses the Amazon S3 Service to store images and tabular data and Amazon EC2 for post-processing before making the data available for decision-making.

\subsubsection{Cleaning}
The data, once in the cloud, is cleaned to ensure integrity, usability and integration with existing GIS files that the city may have of its own street grid. In New York City for example, we use the LION file which is a ``single line representation of New York City streets containing address ranges and other information" [12]. We also use Google Maps' Roads API [13] to ensure consistency of the location information. The Snap to Roads [14] feature improves the accuracy of our GPS traces and allows for a consistent location dataset.

\section{Street Defects Detection Using Computer Vision}
A conventional approach of digital street inspection involves manually annotating individual street images. The approach is not only subjective but also tedious and may introduce other forms of error from mislabeling [15]. To that end, we have developed a computer vision-based framework to automate the detection of various street defects with the purpose to introduce a structured, empirical approach to digitize the current, subjective evaluation criteria of street inspection.

\subsection{Methodology}
Support vector machines or SVM is a widely used technique in supervised learning. Featuring a non-probabilistic binary linear classifier, SVM is well-suited to classifying labeled images in computer vision. Additionally, as an important image processing technique, adaptive thresholding is extensively applied to edge detection, which is extremely helpful to evaluating road textures.
Our methodology combines support vector machines and adaptive thresholding to detect defects in each road image. We have been focusing on crack detection in our efforts thus far. Our current prototype is implemented in Python and leverages the OpenCV library [16].

\subsubsection{Cleaning and Normalization}
Our analysis focuses on the section of road directly behind the vehicle, named the region of interest or ROI; see the left side of Figure 1. The intuition is threefold: one, this region will not generally contain non-road elements, such as sidewalk or other vehicles; two, road elements beyond the upper edge of the ROI will be captured with too few pixels to reliably analyze; and three, the region of the image above the upper edge of the ROI will have been captured in previous images as the vehicle travels down the road.

\begin{figure}[h]
\centering
\includegraphics[width=8cm]{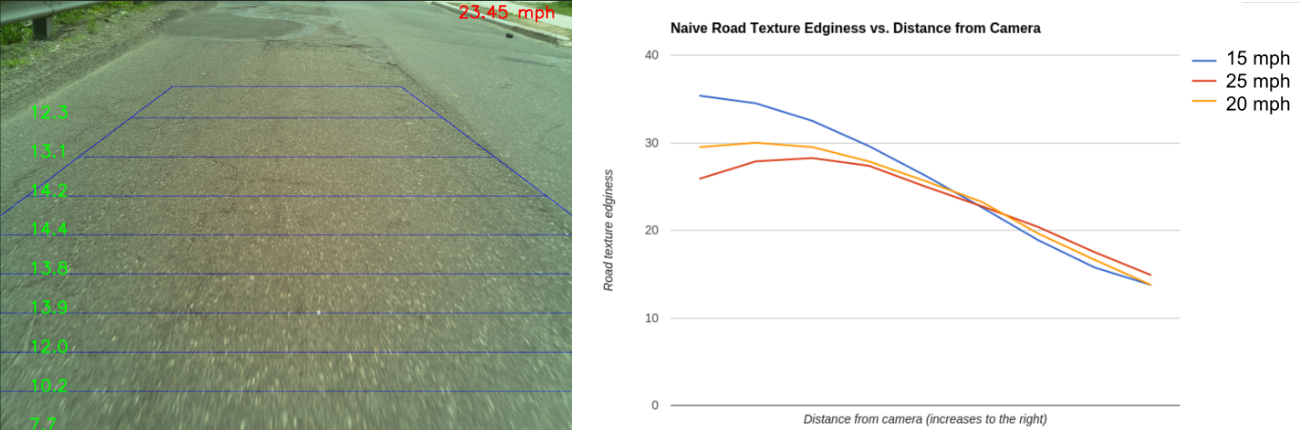}
\caption{Road Texture Edginess versus Distance from Camera.}
\label{road texture edginess}
\end{figure}

\begin{figure*}
  \includegraphics[width=\textwidth]{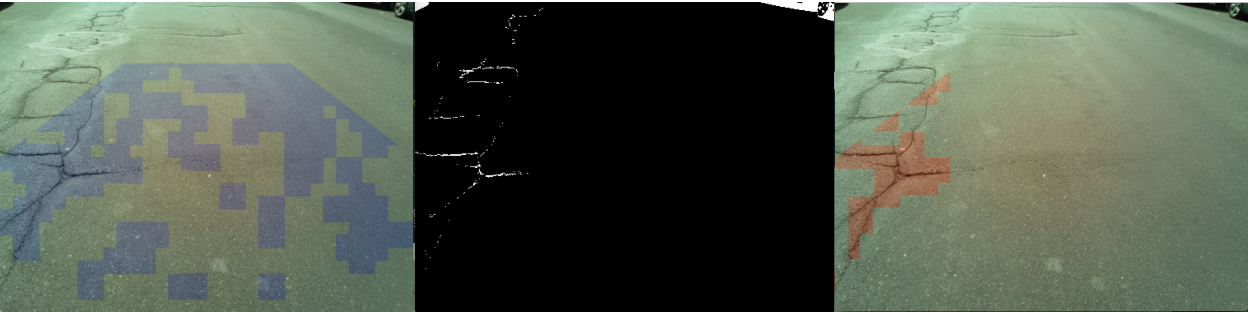}
  \caption{Defect detection result using SVM and adaptive thresholding.}
\end{figure*}

The distance from the camera affects how road texture appears in the image. To evaluate this, we analyzed a number of images of roads with even texture by convolving them with the Sobel operator, which captures image gradients (edges). We split images into a number of horizontal regions, calculated the mean of the Sobel image in each region, and plotted the value as a function of image (a proxy for distance from the camera). We captured images at three different vehicle speeds: 15mph, 20mph, and 25mph; see the right side of Figure 1.

The mean intensity of the Sobel image decreases near the bottom of the image due to motion blur; as expected, this effect is more pronounced at higher vehicle speeds. The mean intensity also decreases with increasing distance from the camera---this is also expected, as higher distances from the camera imply fewer pixels within the ROI.

We therefore normalize each Sobel image with respect to the distance from the camera prior to its use in further processing, in order to reduce the bias in analyzing road texture edginess.

\subsubsection{Man-made Features}
An important consideration is to avoid misclassifying man-made features, such as road markings or manhole covers, as road defects. To this end, we detect these man-made features explicitly and remove them from the ROI prior to looking for defects. We detected white and yellow road markings using simple color- based thresholds followed by morphological operators.

A side benefit of this step is that the quality of road markings can be evaluated. Clearly visible road markings are important to improving street safety for motorists, bicyclists, and pedestrians. Poor quality road markings significantly impede the ability of self-driving cars to function [17]. Making road-marking quality information easily visible and accessible to city administrators will enable stakeholders to direct resources to where they are most needed (see Section 4).

\begin{figure*}
  \includegraphics[width=\textwidth]{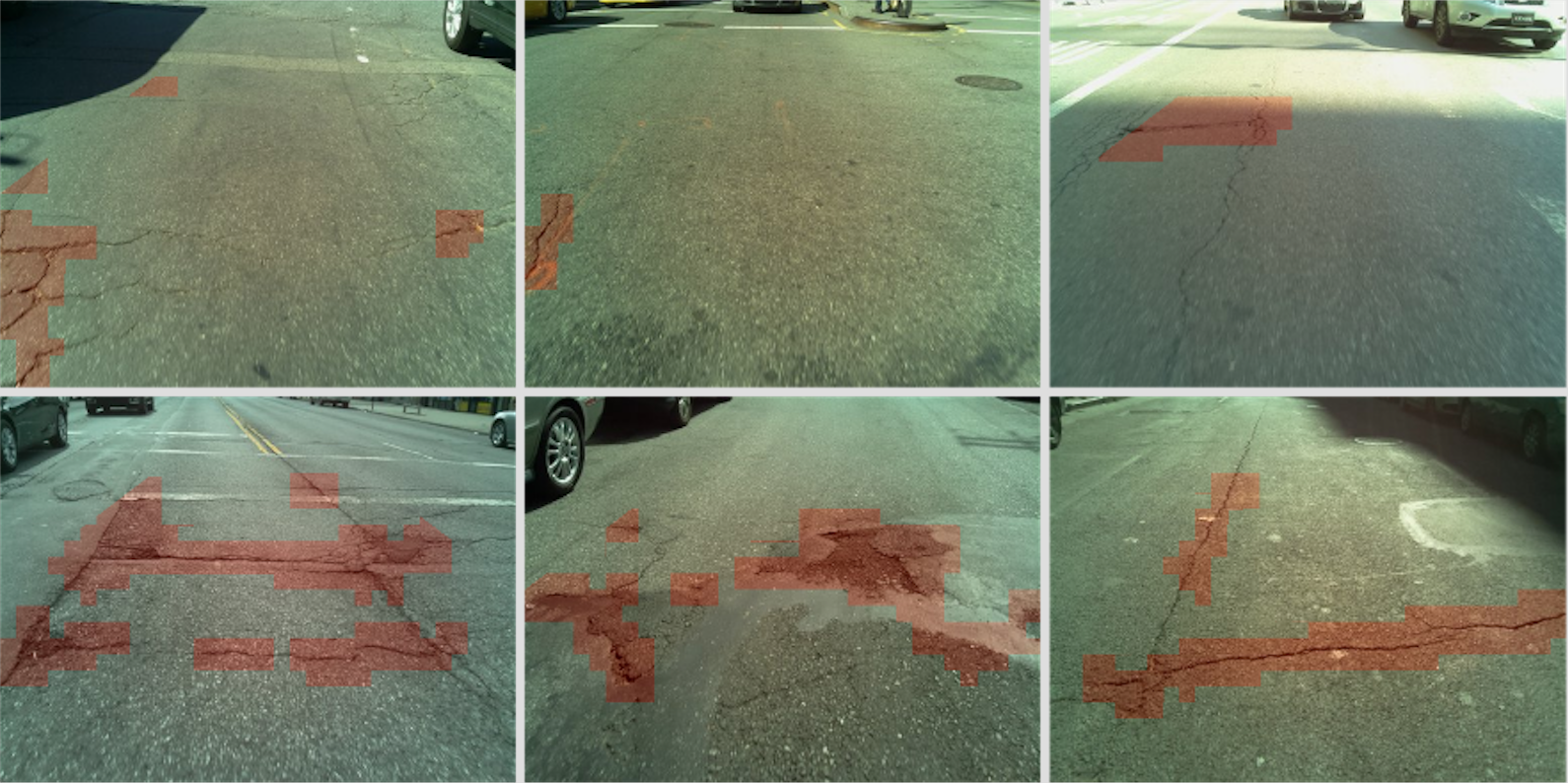}
  \caption{Crack detection results with defect regions highlighted in red.}
\end{figure*}

\subsubsection{Classification Using SVM and Adaptive Thresholding}
Our defect detector combines two individual classifiers: an SVM trained on a manually-annotated subset, and an adaptive threshold pixel-level classifier.

The SVM classifier was trained on a manually-annotated subset of images collected during our prototype deployment in New York City. Example output of the SVM classifier is shown on the left side of Figure 2. As the image shows, the classifier performance is subpar, due largely to the relatively small training dataset. Significant improvements in the overall classification accuracy will be achieved with an improved classifier in this step.

The other subclassifier uses an adaptive threshold. We select pixels that are much darker in the input image than in the blurred image as defect candidates, since this often corresponds to what defects look like in our images, shown in the middle of Figure 2.

In the final step, we combine the outputs of the two classifiers by keeping only regions flagged by the SVM that also have a sufficient number of pixels flagged by the adaptive thresholding. This results in better performance than either of the individual classifiers.

\subsection{Classifier Performance}
Some results are shown in Figure 3. Our prototype detector is fast, simple, and performs well in detecting road defects, particularly considering the limited training dataset size. In addition to improving the performance of the core defect detector, our future work will recognize the defect type and severity, evaluate the quality of road markings, and automatically compute road quality metrics such as the PCI. Other challenging areas include detecting man-made hardware and working reliably in the presence of complicated shadows such as those cast by trees.

\section{User Interface To Digitize Street Inspections}
While computer vision to automate the street inspection process is the end goal of this work, we understand that for this approach to succeed, human expertise is needed to enhancing the efficacy of the defect identification model and also ensures that the architecture can be reused for other purposes such as identifying street assets (furniture, parking signs etc.).

To this end, we have developed a user interface to represent the large quantity of image data that is collected with the intent towards actionable intelligence. In this instance, the User Interface is designed to answer a simple question, ``how do we prioritize all the city's streets for street repair?"

A cornerstone of this interface is to enable stakeholders to engage in a fully virtual inspection process. This is a paradi\-gm shift as current methods of street surveying involve sending individual vehicles and inspectors to the location of the street defect [18]. Enabling a virtual inspection process, we argue would lead to immediate and significant cost-savings as well as environmental benefits simply by getting inspectors and vehicles off the roads and behind digital workstations. Customized training can be delivered to ensure that these city employees who were previously engaged in analog jobs are now retrained to perform digital tasks. To this end, the SQUID approach also unlocks a workforce development opportunity within this section of municipal service delivery.

\subsection{Design}

\begin{figure}[h]
\centering
\includegraphics[width=8cm]{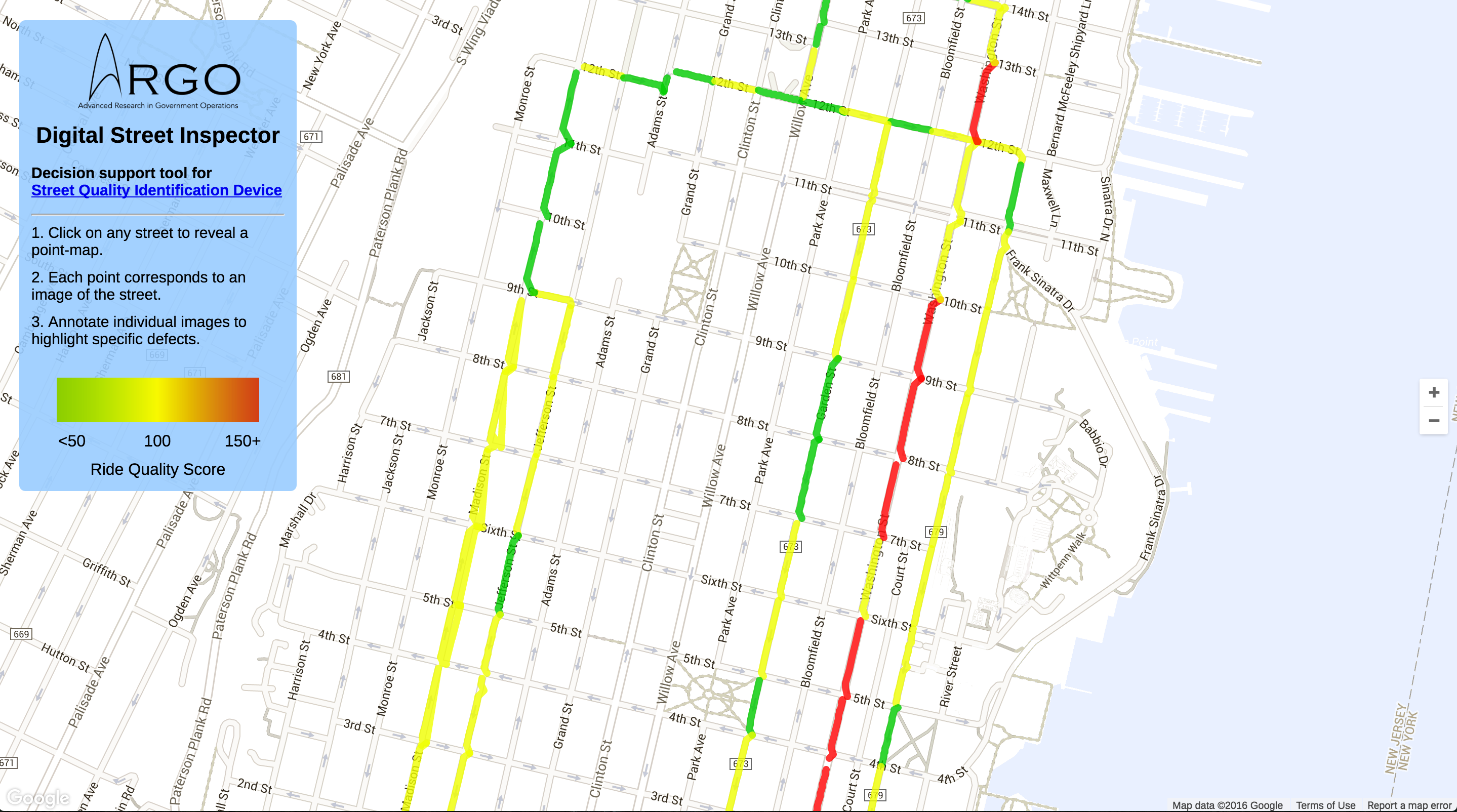}
\caption{SQUID Annotator web dashboard.}
\label{UI main}
\end{figure}

Figure 4 is the main interface of our virtual street inspection tool, implemented using Google Maps API and JavaScript. The demo [19] is designed using data collected in Hoboken, NJ. On the map, data points are displayed along each street we have driven and color-encoded by ride quality score, an index calculated by taking the vector magnitude of acceleration in all three directions (x, y, z). The scores are divided into three groups by value range: 50 and below are encoded in green; 50 to 150 in yellow; and 150 and above in red. By looking at the map, users are able to get a sense of how ride quality score gets distributed at a block level throughout the city. There are about 20 images collected per block, depending on how much time it takes to drive through, usually around 20 seconds.

\subsection{Usability}
The purpose of this User Interface is to make street inspections more efficient by digitizing the entire workflow. By clicking on a data point on the map, the virtual street inspection process begins, allowing users to interact with the UI. SQUID Annotator will display the image taken as well as the detailed data collected at that very instant by our device. In addition to location and time, the data view interface shows the average ride quality of the block as compared to the ride quality of the data point selected, describing the data point at a highly granular level.

\begin{figure}[h]
\centering
\includegraphics[width=8cm]{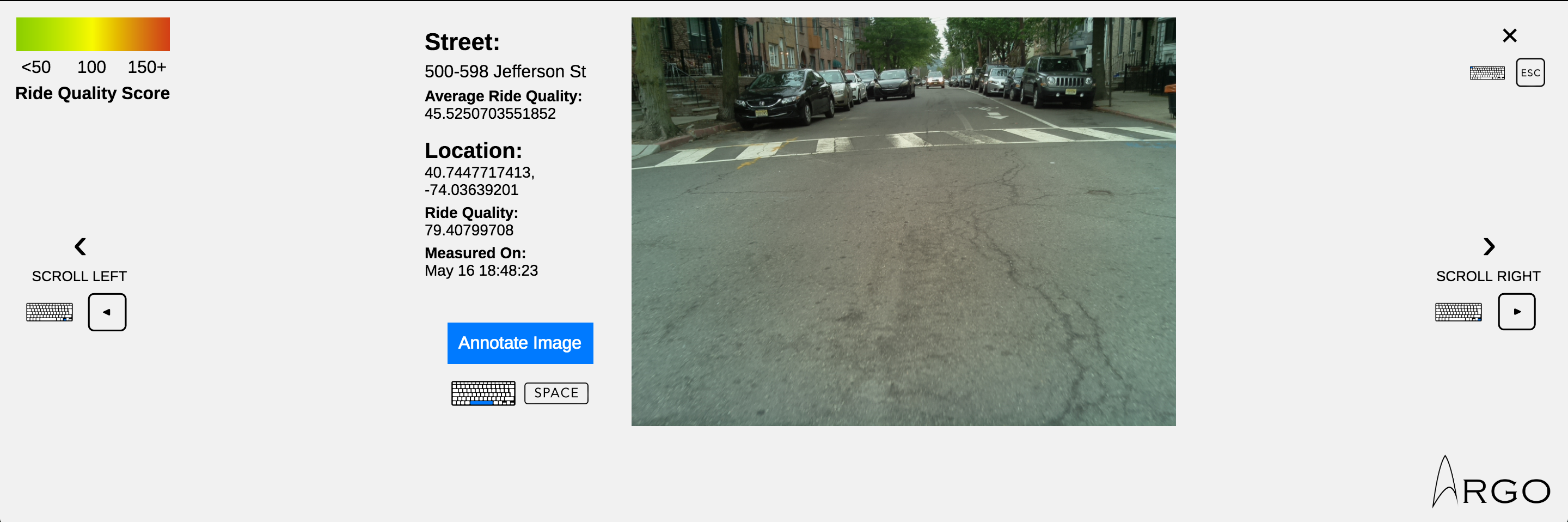}
\caption{SQUID Annotator data view interface.}
\label{UI 1}
\end{figure}

By clicking on the ``Annotate Image" button, users are able to virtually inspect the street. The defect annotation interface provides options for users to select the defect region in the image. After selection, a menu pops up, allowing users to not only record the type of defects but also add a comment to if necessary. Once the annotation is done, a record will be generated at the back end for further processing and analysis.

\begin{figure}[h]
\centering
\includegraphics[width=8cm]{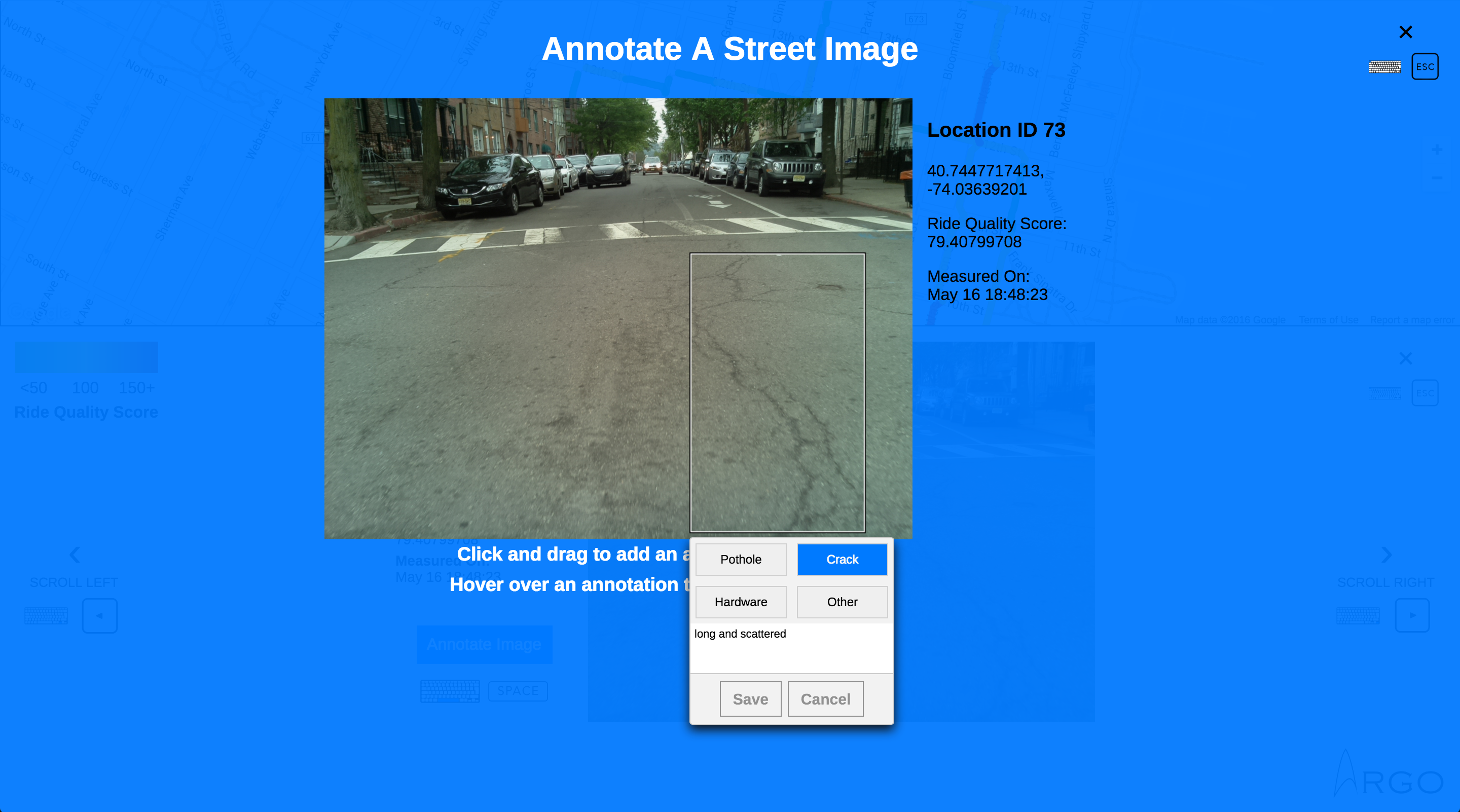}
\caption{SQUID Annotator defect annotation interface.}
\label{UI 2}
\end{figure}

The virtual inspection process is designed to be simple, usable and self-explanatory and over time will merge with the computer vision technique to offer a rich, automated process to digitize street inspection for an entire city.

\section{Deploying SQUID In The City Of Syracuse}
The innovation team at Syracuse, supported by Bloomberg Philanthropies, invited ARGO Labs to prototype SQUID in the city. Beginning on April 14\textsuperscript{th}, 2016 through the 30\textsuperscript{th}, the SQUID device was mounted on 2 city vehicles and collected 110,000 readings and images covering an estimated 538 linear miles of streets. This equated to over half of Syracuse's entire street grid. Our main takeaways from this field experiment included:

\begin{itemize}
  \item Mobile hotspots have a limited capability to upload data and are not suited for continuous image transmission. The preferred approach is to use an asynchronous transmission link using Wifi.
  \item The absence of a structured route plan proved to be a challenge. While the city's department of public works maintains routes in paper-format, a digital approach to route planning would improve the process significantly. Furthermore, the lack of a digital route plan caused instances of ``back-tracking" where the same street segment was surveyed repeatedly.
  \item SQUID was initially mounted on a large truck and then moved to a smaller vehicle. The vibrations from the larger truck introduced noise in our dataset and needed to be normalized.
\end{itemize}

\section{Conclusion and Future Work}
In 2015, Chicago's Office of Inspector General made the following remarks about Chicago's pavement program [20]:

``CDOT is at an auspicious juncture--the tipping point of a paradigm shift from its traditional, reactionary, `worst-first' approach to a comprehensive, proactive pavement management strategy aligned with contemporary best practices that realize the substantial financial benefits of timely, planned preventive maintenance."

We agree with this assessment from the 3rd largest U.S. city and the way to get there is by digitizing municipal street inspections and enabling a 21\textsuperscript{st} century approach to a 20\textsuperscript{th} century problem. On one end, Silicon Valley and private enterprise challenge local governments with visions of autonomous transportation[21,22] while on the other end, the country's road infrastructure that is supposed to support these lofty visions degrades by the day without a sustainable,digital maintenance strategy. Our approach adequately demonstrates how municipalities can indeed be smart by meaningfully combining technology with a mission to serve the public and deliver core services efficiently, transparently, and at scale.

SQUID's future work involves several improvements to the overall work flow around data collection, storage and eventual analysis. On the collection end, we have begun prototyping a light-weight mobile application that does not require a dedicated hardware solution. In this deployment, a municipal employee may need only to mount the smart phone to the rear of the vehicle with a clear view of the street to enable scalable data collection.

Moreover, the base models of today's smart phones are equipped with all the necessary sensors and optical technology needed to deliver a low-cost, high-quality digital street quality assessment for the entire city.

To date we have collected approximately 300,000 images of streets that forms our initial training data set. We are confident that with similar deployments such as in Syracuse, our computer vision model to detect street defects and pavement markings will only improve with time and these improvements can be deployed to all participants. 

A goal is to also afford a structured and longitudinal data collection process on a single test site, over a longer period  of time. This would begin unlocking a preventative paradigm for street maintenance that is truly unprecedented and could lead to exponential improvements and cost-savings.

Finally, we are also working on an automated route-planning service that will allow a municipal employee to create a programmatic driving plan so that the entire street grid can be driven in a structured manner. This would enable us to deliver this technology to non-municipal, private fleets as well.
\newline

\section{Acknowledgments}
We would like to thank the city of Syracuse; their forward looking mayor, Stephanie Miner, the Syracuse I-team lead by Andrew Maxwell, Sam Edelstein, Adria Finch, and Addison Spears for providing us the opportunity to prototype SQUID and engaging with the city's department of public works who were very forthcoming and co-operative. We would also like to thank the Knight Foundation for supporting this work through the Prototype fund without which these efforts and advances would not have been possible.
\newline
\newline
\newline
\newline

\nocite{*}
\bibliographystyle{unsrt}
\bibliography{references}

\end{document}